\documentclass[a4paper,11pt]{article}
\usepackage{pos}
\usepackage{graphicx} 
\graphicspath{
	{operators/},{figures/}} 
\usepackage[font=small,labelfont=bf, figurename=Fig.]{caption} 
\usepackage{amsmath} 
\usepackage{color, colortbl}
\usepackage{diagbox}
\usepackage{subcaption}
\usepackage[section]{placeins}
\usepackage{ tabularx}
\newcommand{\Tr}{\text{Tr}}
\newcommand{\im}{\mathrm{i}}
\newcommand{\qu}[1]{\overset{\circlearrowleft}{q}_{#1}}
\newcommand{\aqu}[1]{\overset{\circlearrowleft}{\bar{q}}_{#1}}
\newcommand{\quc}[1]{\overset{\circlearrowright}{q}_{#1}}
\newcommand{\aquc}[1]{\overset{\circlearrowright}{\bar{q}}_{#1}}

\title{Eight spectra of very excited flux tubes in SU(3) gauge theory}

\author*[a]{Pedro Bicudo}
\author[a]{Alireza Sharifian}
\author[a]{ Nuno Cardoso}

\affiliation[a]{CeFEMA, Departamento de Física, Instituto Superior Técnico,\\
 Av. Rovisco Pais, 1049-001 Lisboa, Portugal}

\emailAdd{ bicudo@tecnico.ulisboa.pt}
\emailAdd{ alireza.sharifian@tecnico.ulisboa.pt}
\emailAdd{nuno.cardoso@tecnico.ulisboa.pt}

\abstract{We compute the spectra of flux tubes formed between a static quark antiquark pair up to a significant number of excitations and for eight symmetries of the flux tubes, up to  $\Delta_u$, using pure  $SU(3)$ gauge lattice QCD in 3+1 dimensions. To accomplish this goal, we use a large set of appropriate operators, an anisotropic tadpole improved action, smearing techniques, and solve a generalized eigenvalue problem. Moreover, we compare our results with the Nambu-Goto string model to evaluate possible tensions which could be a signal for novel phenomena.}

\FullConference{%
The 39th International Symposium on Lattice Field Theory\\
  8th – 13th August, 2022\\
  Hörsaalzentrum Poppelsdorf,  University of Bonn, Bonn, Germany.
}

\newcommand{\spectraCaption}{lines show Eq. \eqref{eq:modified_nambu_goto} fitted to the simulation data and the width of the highlighted area around them corresponds to the error of the fit. $\lambda=0,\ 1,\ \ldots$ show the ground state, first excitation, and so on, depicted with the same color as the corresponding quantum number $N$. \textit{Fit for  $N:N_i\to N_f$} indicates the values of $N$ included in the fit. Gray highlighted areas show the charge separation $R$ included in the fit. }
\begin{document}
\maketitle

\section{Introduction}\label{sec:introduction}
 As gluons, force carries of strong forces, have color charges, the gluonic fields are squeezed in the vacuum and form a flux tube. This is in contrast to the electromagnetic fields which spread out in the space. The dominant behavior of flux tubes are string-like. A confirmation for the string-like behavior is the Regge trajectories observed in hadron spectra. The string theories also predict a linear potential between quarks which is confining and reproduces correctly the confinement of quarks inside hadrons.

Quantization of a relativistic string leads to a tower of excitations, however different theoretical models exist for the excitations of hadrons, such as bag models for different sorts of hadrons, or a few-body potentials for mesons, baryons or hybrids. Therefore, a first principle computation is important to test these models and search for novel phenomena. Numerous lattice QCD calculations have been devoted to study the excitation of the flux tube \cite{  Juge_2003,Capitani:2019}, however, they only succeeded to compute a small number of excitations, up to two excitations for the most amenable symmetries of the flux tube. In this work, we continue our previous study of the $\Sigma_g^+$ spectrum \cite{Bicudo:2021tsc} and compute a significant number of excitations for other symmetries of the flux tube. We also compare our results with the Nambu-Goto string model.

The Nambu-Goto string model is defined by the action
\begin{align}
	S=-\sigma \int \mathrm{d}^2\Sigma, 
\end{align}
where $\sigma$ is the string tension and $\Sigma$ is the surface of the worldsheet swept by the string. The energy of an open relativistic string with length $R$ and fixed ends is obtained as 
\begin{align}
	&V(R)=\sqrt{\sigma^2 R^2+2\pi\sigma (N-(D-2)/24)},
	\label{eq:nambu_goto}
\end{align}
where $N$ is the quantum number for string vibrations and $D$ is the dimension of space time. This expression is known as the Arvis potential \cite{ARVIS1983106}.

There are three constants of motion whose eigenvalues are used to label the quantum state of the flux tube:

\begin{minipage}{\columnwidth}
	\begin{minipage}{0.65\columnwidth}
		\begin{itemize}
			\item $z$-component of angular momentum $\Lambda=0,1,2,3,\ldots$, they are typically denoted by greek letters  $\Sigma,\ \Pi,\ \Delta,\ \Phi, \ldots$, respectively
			\item Combination of the charge conjugation and spatial inversion with respect to the mid point of the charge axis operators  $\mathcal{P}o\mathcal{C}$, its  eigenvalues  $\eta=1,-1$, they are denoted by $g,\ u$, respectively.
			\item For $\Sigma (\Lambda=0)$, reflection with respect to a plane containing the charge axis, $\mathcal{P}_x$, with  eigenvalues $\epsilon=+,-$.
		\end{itemize}
	\end{minipage}\hfill
	\begin{minipage}{0.35\columnwidth}
		\centering
		\includegraphics[scale=1.5]{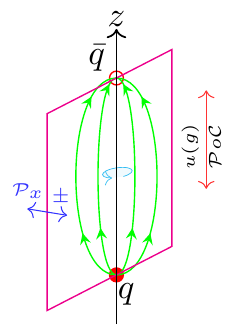}
		\captionof{figure}{Flux tube symmetries.}
	\end{minipage}
\end{minipage}
Therefore, all possible quantum numbers are:
$\Sigma_g^+$, $\Sigma_g^-$, $\Sigma_u^+$,$\Sigma_u^-$,$\Pi_g$,$\Pi_u$, $\Delta_u$,$\Delta_g$, $\ldots$.
\section{Tadpole improved action}
\begin{figure}[hb]  	
	\centering
	\includegraphics[scale=1.2]{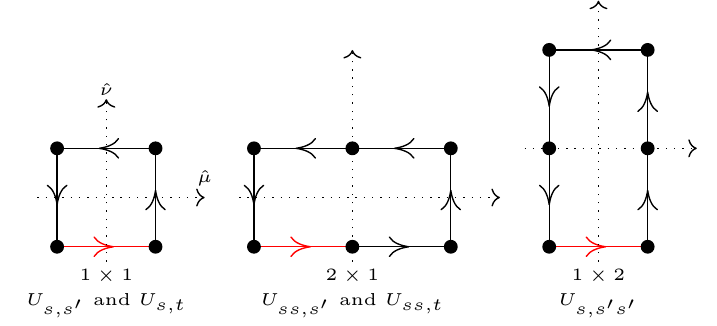}
	\caption{Closed loops $U_c$ are in the anisotropic tadpole improved action of Eq. \eqref{eq:S2}. The redlink is updated in each step of the Monte Carlo method. \label{fig:improved_action}}
\end{figure}

In this work, we use the tadpole improved action on anisotropic lattices which is defined as,  
\begin{align}
	S_\text{II} &= \beta \big( \frac{1}{\xi}     \sum_{x,s>s'}  \big[\frac{5 W_{s,s'}}{3 u_s^4} - \frac{W_{ss,s'}+W_{s,s's'}}{12 u_s^6} \big]
	+  \xi \sum_{x,s} \big[\frac{4 W_{s,t}}{3 u_s^2 u_t^2} - \frac{W_{ss,t}}{12 u_s^4 u_t^2}  \big]\big), \label{eq:S2}
\end{align}
where $\beta=6/g^2$ is the so-called inverse coupling and $\xi$ is the bare anisotropic factor defined as the ratio of spatial lattice spacing to temporal lattice spacing $(a_s/a_t)$. Furthermore, $W_c={1 \over 3}\sum_c \Re e\  \Tr [1-U_c]$ where $U_c$ are the closed loops shown in Fig. \eqref{fig:improved_action}.  Tadpole improvement factors   $u_s = \langle {1 \over 3}\Re e\  \Tr [U_{s,s'}] \rangle^{1/4}$ and $u_t= \langle {1 \over 3} \Re e\ \Tr [U_{s,t}] \rangle^{1/2}/u_s$. This action has a smaller discretization error than the standard Wilson actions. Anisotropic actions $(\xi>1)$ have more time slices in plateaux compared to isotropic ones ($\xi=1$) as well. As a result, we obtain a better estimation for the effective mass \cite{Morningstar_1997}. 
In Table. \eqref{table:configuration_properties}, we listed the properties of pure $SU(3)$  gauge configurations used in this work.

\begin{table}
	\centering
	{\scriptsize
		\begin{tabular}{cccccccc|c|c}
			\hline\hline
			$\beta$&$\xi$&$\xi_R$&volume& $u_s$ &$u_t$& $a_s\sqrt{\sigma}$&$a_t\sqrt{\sigma}$&Smearing (space, time)& No. of configs\\
			\hline
			4&4&3.6266(32)&$24\times96$&0.82006&1&0.3043(3)&0.0839(1)&
			$Stout_{0.15}^{20}$ ,Multihit(100)& 1060\\
			\hline\hline
		\end{tabular}
	}
	\caption{Properties of the pure $SU(3)$ gauge configurations generated by the action of Eq. \eqref{eq:S2}. $\xi_R$ is the renormalized anisotropic factor \cite{Bicudo:2021tsc}.\label{table:configuration_properties}}
\end{table}

\section{Obtaining the excitations}
\begin{figure}[ht]
	\centering
	\includegraphics[scale=1.15]{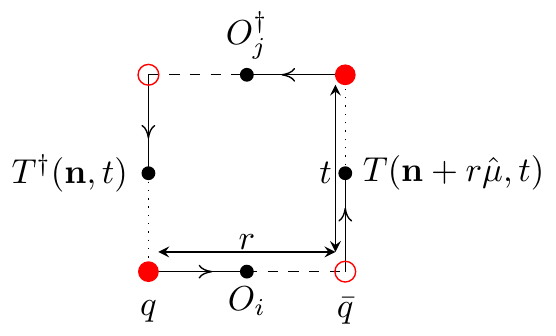}
	\caption{ A closed loop corresponds to the entry $\mathcal{C}_{i,j}(r,t)$ of the correlation Wilson matrix.\label{fig:wilson_loop}}
\end{figure}

To compute the spectra of the flux tube, we first compute the Wilson correlation matrix $\mathcal{C}(r, t)$. The entry $\mathcal{C}_{i,j}(r, t)$ of the Wilson correlation matrix is the expectation value of spatial-temporal closed loops, Fig. \eqref{fig:wilson_loop}, whose spatial sides are replaced with operators $O_i$ and $O_j$ having identical symmetry to the flux tube of interest. 
Afterwards, we find generalized eigenvalues $\lambda$ of the Wilson correlation matrix, 
\begin{align}
	\mathcal{C}(r, t)\vec{\nu}_n=\lambda_n(r, t) \mathcal{C}(r, t_0) \vec{\nu}_n, 
\end{align}
where we set $t_0=0$. Consequently, we obtain a set of time dependent eigenvalues $\lambda_n(t)$ for each $r$. Then, we order the eigenvalues and plot the effective mass defined as
\begin{align}
	E_i(r)=\ln \frac{ \lambda_i(r, t)}{\lambda_i(r, t+1)}.
\end{align}
The plateau in the effective mass plot corresponds to the energy $E_i(r)$.

Note that plateaux usually appear in large time values where contamination from excited states are suppressed; however, effective mass plots for large $t$ are very noisy. To circumvent this problem, we apply stout and multihit smearing for spatial and temporal gauge links, respectively. 
\section {Operators and their spectra}
In this part, we introduce the operators used for different symmetries of the flux tube. These operators are selected such that they lead to a smaller energy for the ground state of each symmetry and lead to smaller noises for the excited states as well. Moreover, we opt out the operators that lead to degeneracies in the spectra. The idea is to select operators with different distances $l$ from the charge axis to sweep the width of the flux tube as much as possible. Because the spatial length of our lattice is 24 lattice spacings and it has periodic boundary conditions, we select the value of $l$ up to 12 lattice links.
\subsection{Analysis of our results}
In the following sections, we show the operators and obtained spectra  for different symmetries of the flux tube, up to $\Delta_u$, as a function of the charge distance $R$ .To set the scale, we fit the ground state of $\Sigma_g^+$ to $V(r)=V_0+\sigma r$. Then,  we use the value of $\sigma$ to present our results in the string tension unit ($\sqrt{\sigma}$).

When we compare our result with the Nambu-Goto spectrum Eq. \eqref{eq:S2}, we observe excited states departure from the Nambu-Goto model. To quantify this tension, we fit our results to
\begin{align}
&V_1(R)=\sigma R\sqrt{1+\frac{2\pi}{\sigma_2 R^2} (N-(D-2)/24)},
\label{eq:modified_nambu_goto}
\end{align}
where $\sigma_2< 1$ corresponds to a larger gap between excited states than in the Nambu-Goto string model. This deviation can be interpreted as the existence of constituent gluons in the more excited states. 

 We fit the data of all excitations with clear signals to Eq. \eqref{eq:modified_nambu_goto}. Notice that the corresponding excitation numbers $N$ are fixed and we exclude the data for small charge separations $r$.  \textcolor{black}{The values of $N$ for each symmetry derived from Ref. \cite{JUGE_2004}.}  
\subsection{Operators for $\Sigma_g^+$, $\Pi_u^+$, $\Delta_g^+$ symmetries and the spectra}
\begin{figure}[ht]
		\centering
	\begin{minipage}{0.5\textwidth}
		\centering
		\includegraphics[scale=0.9]{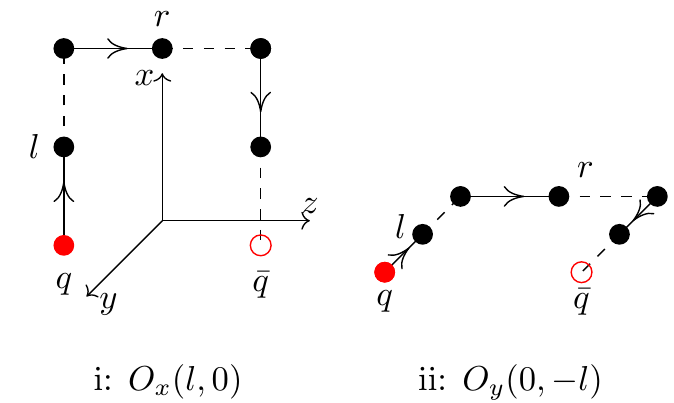}
\subcaption{we vary $l$  from 0 (for $\Sigma_g^+$) or 1 to 12 to construct 13 or 12 operators. \label{op:Oi}}
\end{minipage}\hfill
	\begin{minipage}{0.5\columnwidth}
	\includegraphics[scale=0.8]{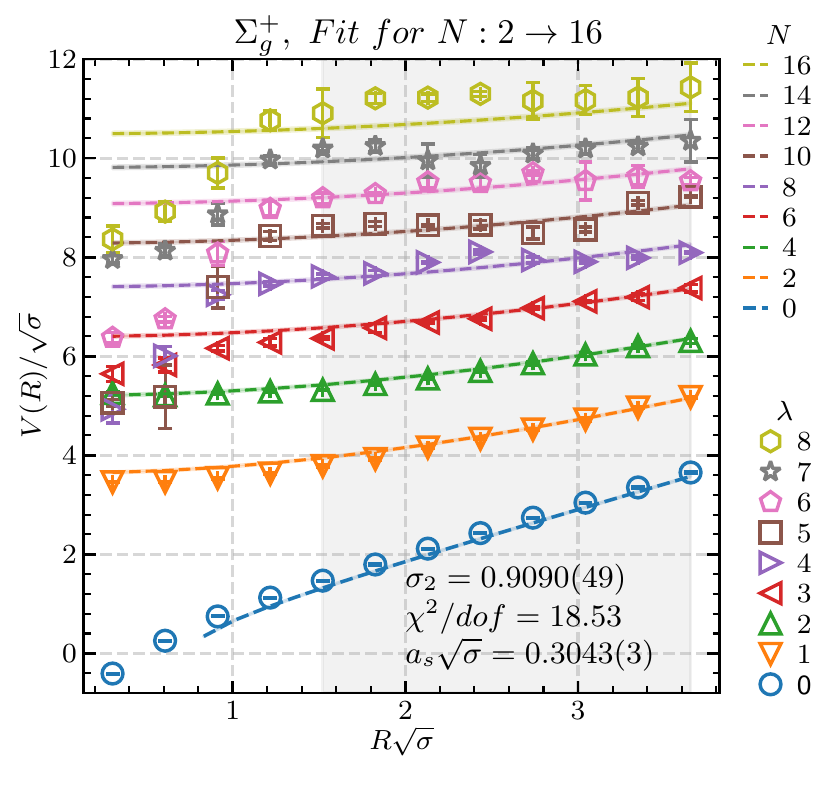}
	\subcaption{$\Sigma_g^+$ spectrum, data derived from \cite{Bicudo:2021tsc}.\label{fig:sigma_gp_spectra}}
\end{minipage}\vfil	
	\begin{minipage}{0.5\columnwidth}
	\includegraphics[scale=0.8]{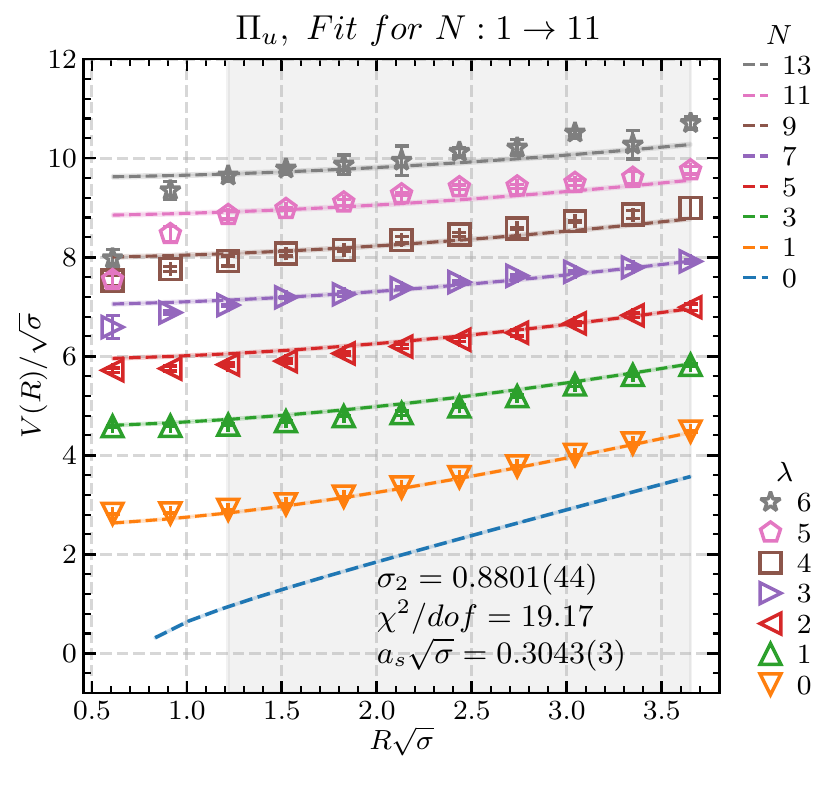}
	\subcaption{$\Pi_u$ spectrum.\label{fig:pi_u_spectra}}
\end{minipage}\hfil
	\begin{minipage}{0.5\columnwidth}
	\includegraphics[scale=0.8]{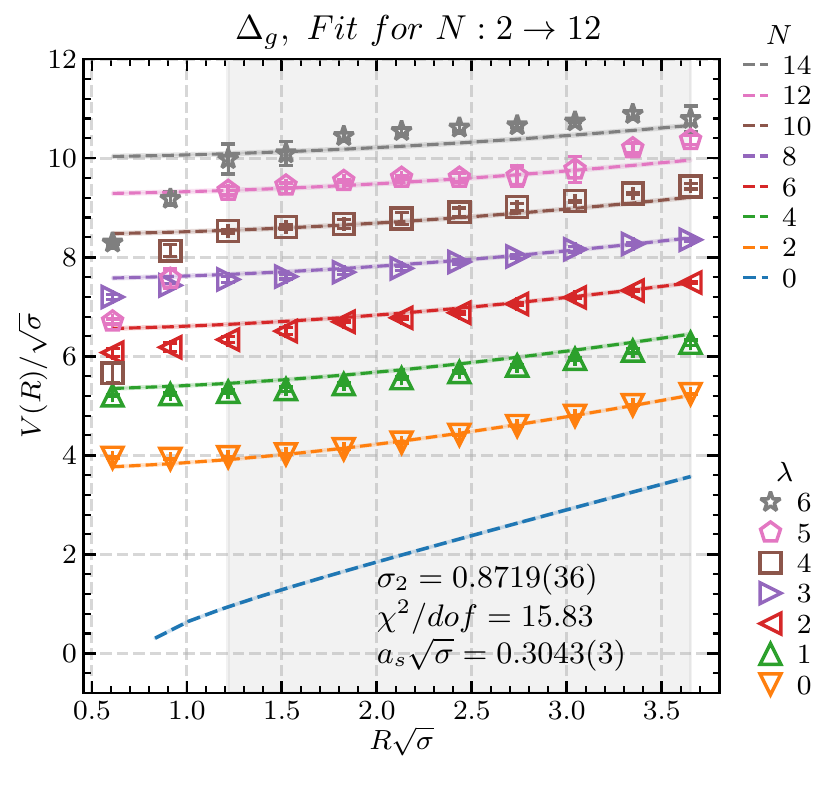}
	\subcaption{$\Delta_g$ spectrum.\label{fig:delta_g_spectra}}
\end{minipage}
\caption{$\Sigma_g^+$, $\Pi_u^+$, and $\Delta_g^+$ spectra. Dashed \spectraCaption }
\end{figure}

We use suboperators shown in Fig. \eqref{op:Oi} to construct operator with $\Sigma_g^+$, $\Pi_u^+$, and $\Delta_g^+$ symmetries. Consequently, the operators are written as
\begin{align}
&O^{\Sigma_g^+}(l, 0)=\frac{1}{2}\big(O_x(l,0)+O_y(l,0)+O_x(-l,0)+O_y(0,-l)\big),\\
&O^{\Pi_u^+}(l, 0)=\frac{1}{2}\left(O_x(l,0)+ \im O_y(l,0)-O_x(-l,0)-\im O_y(0,-l)\right),\\
&O^{\Delta_g^+}(l, 0)=\frac{1}{2}\left(O_x(l,0)-O_y(l,0)+O_x(-l,0)-O_y(0,-l)\right).
\end{align}
The spectra obtained using these operators are shown in Figs. \eqref{fig:sigma_gp_spectra}, \eqref{fig:pi_u_spectra}, and \eqref{fig:delta_g_spectra}, respectively.

\begin{figure}[!th]
	\begin{minipage}{0.5\textwidth}
		\centering
		\includegraphics[scale=0.85]{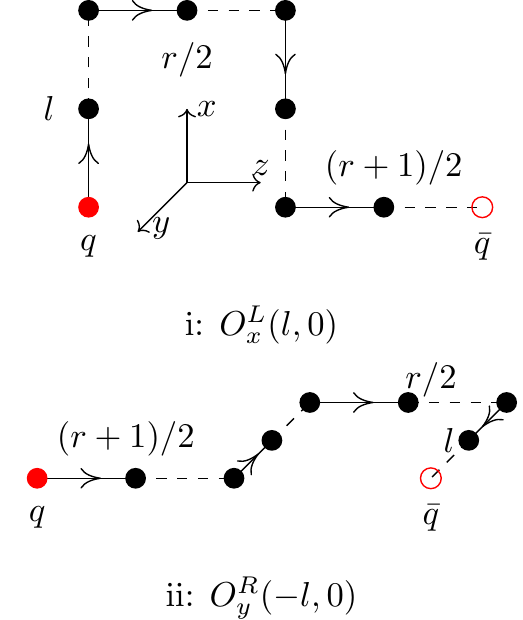}
		\subcaption{We vary $l$ from $1$ to $12$ to build $12$ operators.\label{op:halfline}}
	\end{minipage}\hfill
	\begin{minipage}{0.5\columnwidth}
		\includegraphics[scale=0.8]{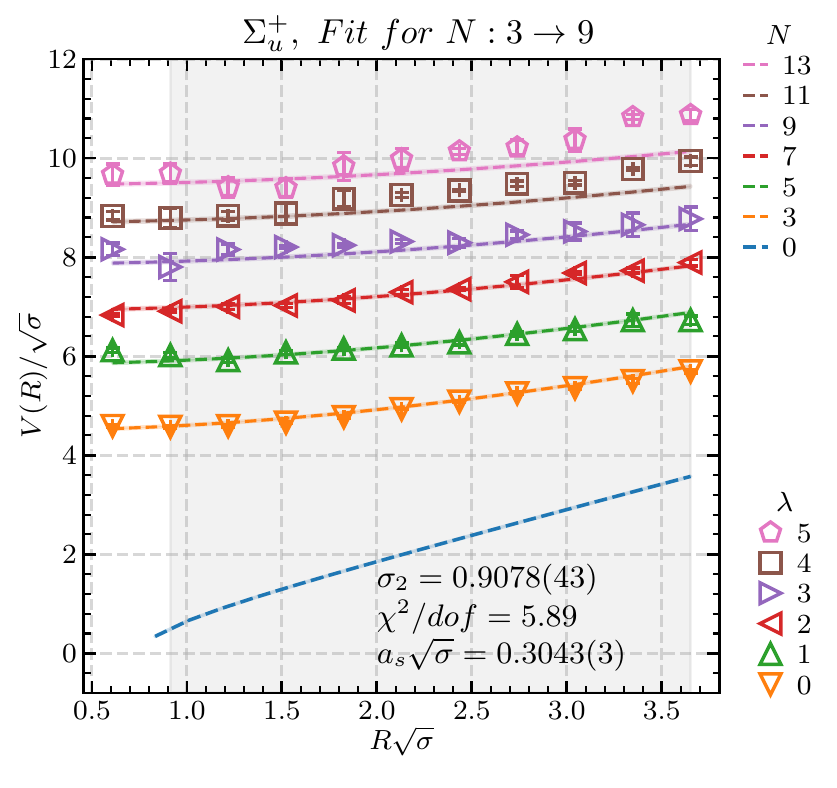}
		\subcaption{$\Sigma_u^+$ spectrum \label{fig:sigma_up_spectra}}
	\end{minipage}\vfil
	\begin{minipage}{0.5\columnwidth}
		\includegraphics[scale=0.8]{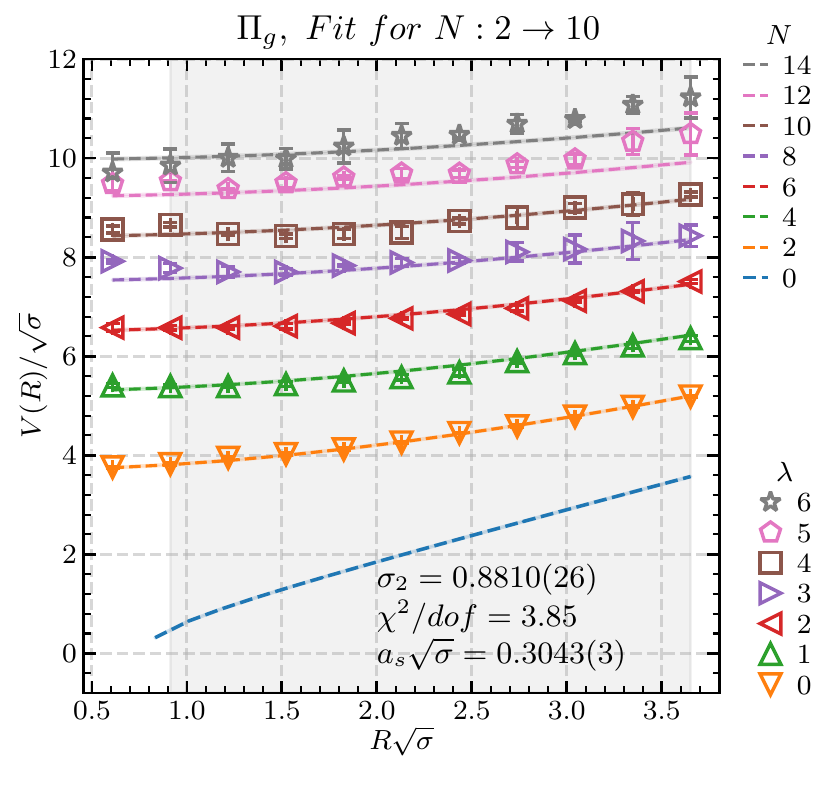}
		\subcaption{$\Pi_g$ spectrum.\label{fig:pi_g_spectra}}
	\end{minipage}\hfill
	\begin{minipage}{0.5\columnwidth}
		\includegraphics[scale=0.8]{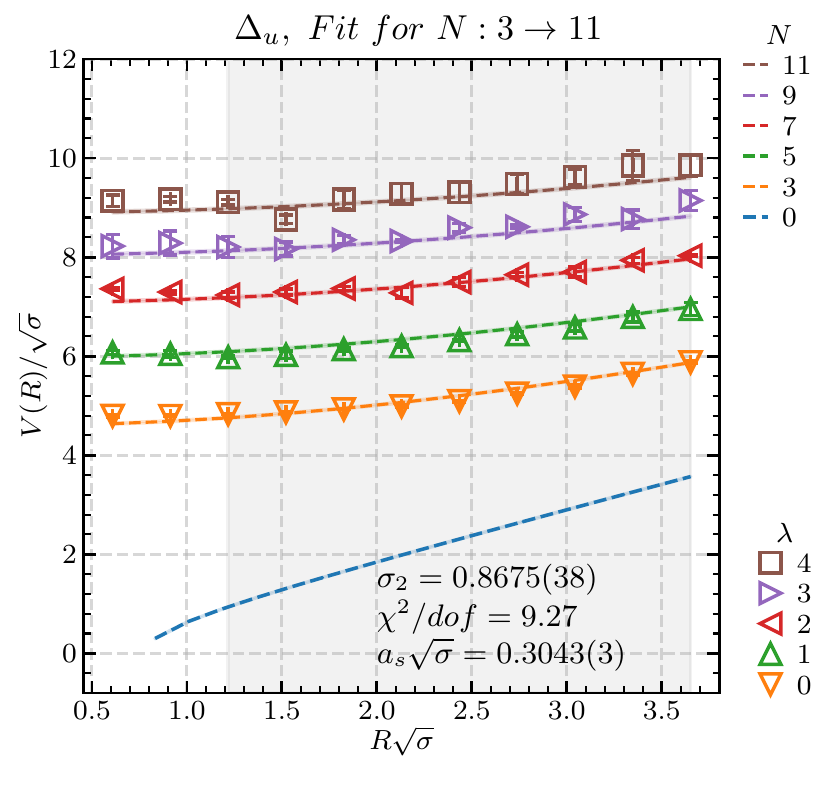}
		\subcaption{$\Delta_u$ spectrum.\label{fig:delta_u_spectra}}
	\end{minipage}
	\caption{$\Sigma_u^+$, $\Pi_g$, and $\Delta_u$ spectra. Dashed \spectraCaption}
\end{figure}
\subsection{Operators for $\Sigma_u^+$, $\Pi_g^+$, $\Delta_u^+$ symmetries and the spectra}
For $\Sigma_u^+$, $\Pi_g^+$, $\Delta_u^+$  symmetries of the flux tube, we use half-line suboperator shown in Fig. \eqref{op:halfline}. The explicit formulas of these symmetries are obtained as
 \begin{align}
O^{\Sigma_u^+}_{1/2}(l,0)=\frac{1}{2\sqrt{2}}\bigg(&O^L_x(l,0)+O^L_y(l,0)+O^L_x(-l,0)+O^L_y(-l,0)\nonumber\\
&-O_x^R(-l, 0)-O^R_y(-l,0)-O_x^R(l, 0)-O_y^R(l, 0)\bigg),
\end{align}\begin{align}
O^{\Pi_g^+}_{1/2}(l,0)=\frac{1}{2\sqrt{2}}\bigg(&
O_x^L(l,0)+\im O_y^L(l,0)-O_x^L(-l,0)-\im O_y^L(-l,0)\nonumber\\
&-\big[ O_x^R(l,0)+\im O_y^R(l,0)-O_x^R(-l,0)-\im O_y^R(-l,0)\big]
\bigg),\\
O^{\Delta_u^+}_{1/2}(l,0)=\frac{1}{2\sqrt{2}}\bigg(&
O_x^L(l,0)- O_y^L(l,0)+O_x^L(-l,0)- O_y^L(-l,0)\nonumber\\
&-\big[ O_x^R(l,0)- O_y^L(l,0)+O_x^R(-l,0)- O_y^R(-l,0)\big]
\bigg). 
\end{align}
In Figs. \eqref{fig:sigma_up_spectra}, \eqref{fig:pi_g_spectra}, and \eqref{fig:delta_u_spectra}, we show the spectra obtained using these operators.
\subsection{Operators for $\Sigma_g^-$ symmetry and the spectrum}
\begin{figure}[!ht]
	\centering
	\begin{minipage}{0.5\textwidth}
			\vspace*{1.5cm}
		\includegraphics[scale=1]{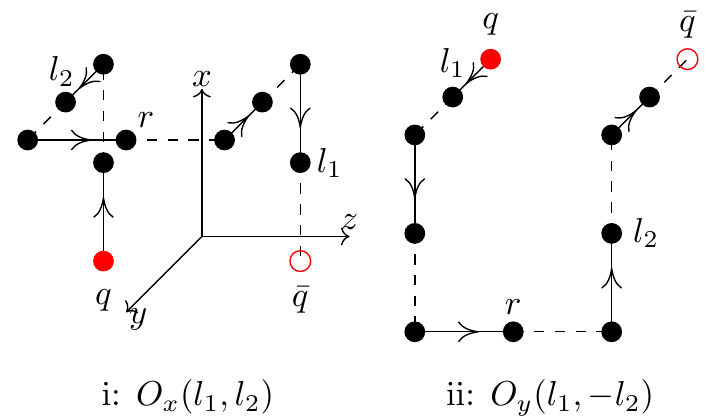}
			\vspace*{0.5cm}
		\subcaption{	We fix $l_1=1$ and let $l_2$ vary from 1 to 12, so we end up with 12 operators.\label{fig:sigma_gm_operator}}
	\end{minipage}\hfil
	\begin{minipage}{0.5\columnwidth}
		\includegraphics[scale=0.8]{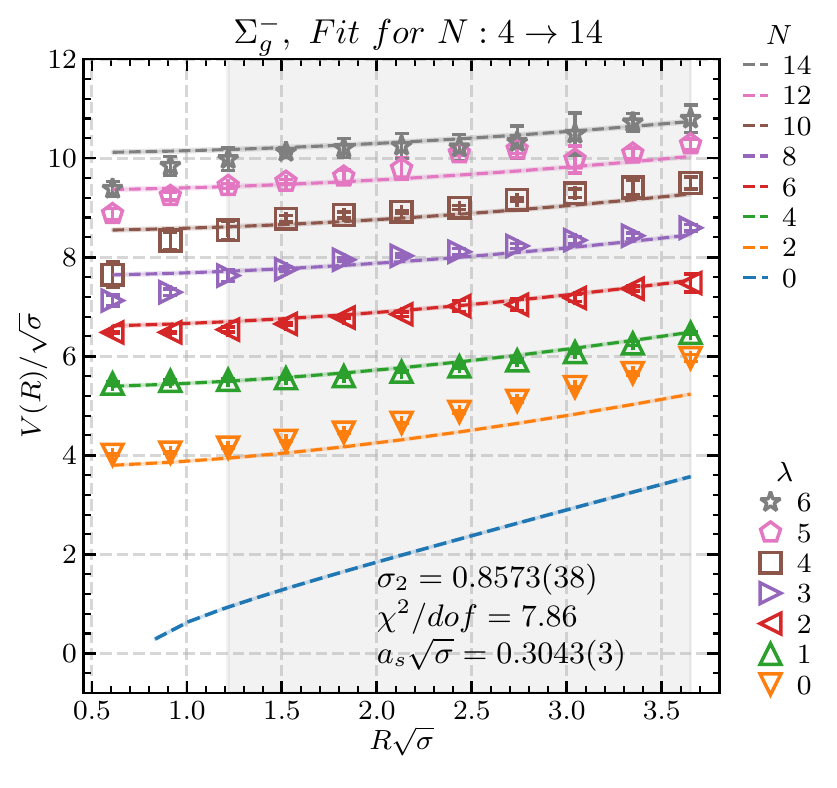}
		\subcaption{$\Sigma_g^-$ spectrum. \label{fig:sigma_gm_spectra}}
	\end{minipage}
	\caption{ $\Sigma_g^-$ suboperators and the spectrum. Dashed \spectraCaption}
\end{figure}

To construct operators with $\Sigma_g^-$ symmetry, we use the suboperators shown in Fig. \eqref{fig:sigma_gm_operator}. There are several options for the values of $l_1$ and $l_2$ in Eq. \eqref{eq:sigma_gm_operator}. We choose values for $l_2$ that range from 1 to 12 while fixing the value of $l_1=1$.
\begin{align}
O^{\Sigma_g^-}(l_1, l_2)&=\frac{1}{2\sqrt{2}}\bigg(O_x(l_1,l_2)-O_x(l_1,-l_2)+O_x(-l_1,-l_2)-O_x(-l_1,l_2)+O_y(l_1,-l_2)\nonumber\\
&-O_y(-l_1,-l_2)+O_y(-l_1,l_2)-O_y(l_1,l_2)\bigg)\label{eq:sigma_gm_operator}
\end{align}

\textcolor{black}{In Fig. \eqref{fig:sigma_gm_spectra}, we could  fit  the ansatz of Eq. \eqref{eq:modified_nambu_goto} to our data properly only if we exclude the ground state data (orange points) from the fit and assign $N=4$ to the first excitation (green points). This is in contrast to the prediction of string models  which assign $N=4$ to the ground state of $\Sigma_g^-$ \cite{JUGE_2004}. Therefore, this state might correspond to another particle, say an axion \cite{Dubovsky:2013gi,Athenodorou:2021vkw}. }
\subsection{The operator for $\Sigma_u^-$ symmetry and the spectrum.}
\begin{figure}[ht]
\begin{minipage}{\textwidth}
	\centering
	\includegraphics[scale=0.7]{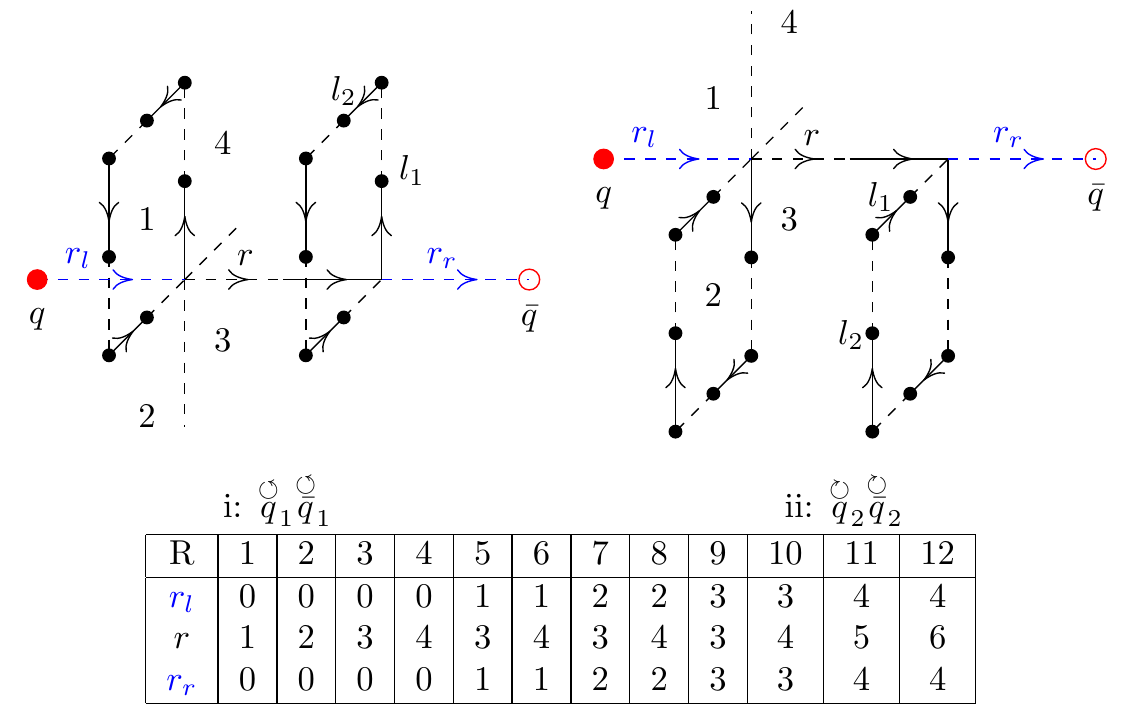}
	\subcaption{We select $l_1=l_2$ and let $l_1$ vary from 1 to 12 to build 12 operators.  The values of $r_l$ and $r_r$ derived from Ref. \cite{Capitani:2019}. \label{fig:op_sigma_um}}
\end{minipage}\vfil
\begin{minipage}{\textwidth}
		\centering
	\includegraphics[scale=0.75]{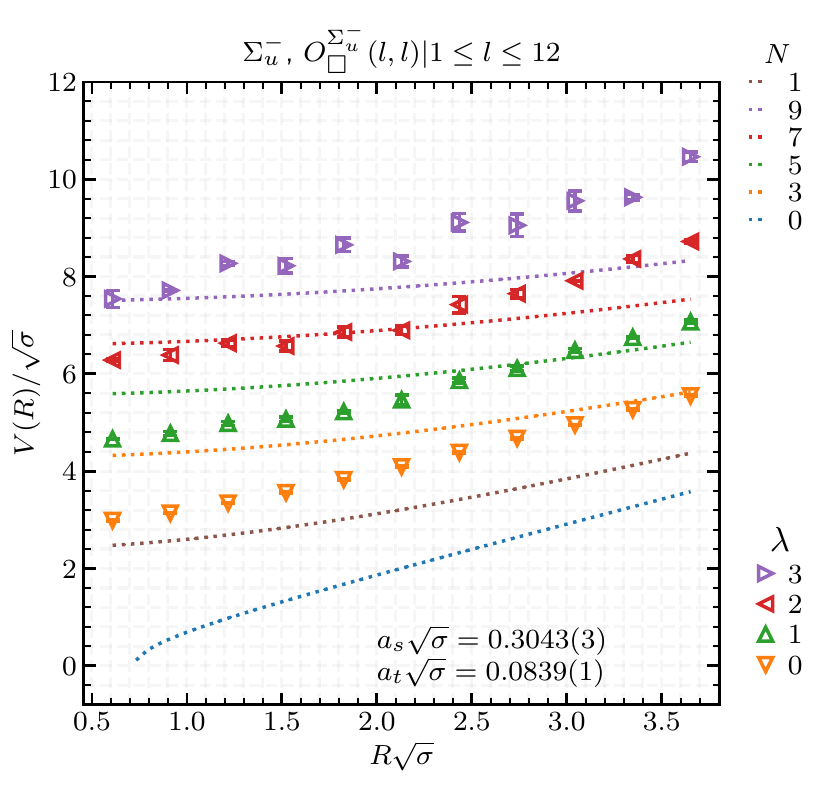}
	\subcaption{$\Sigma_u^-$ spectrum. \label{fig:sigma_um_spectra}}
\end{minipage}
\caption{$\Sigma_u^-$  suboperators and the spectrum, dotted lines show the Nambu-Goto model spectrum, Eq. \eqref{eq:nambu_goto}. $\lambda=0,\ 1,\ 2, \ldots$ denote the ground state, first excitation, and so on,  depicted with the same color as the corresponding quantum number $N$.}
\end{figure}

For $\Sigma_u^-$ symmetry, we use loop like suboperators shown in Fig. \eqref{fig:op_sigma_um}. So we obtain 
\begin{align}
O^{\Sigma_u^-}_{\square}(l,l)
&=\frac{1}{2\sqrt{2}}\bigg(\qu{1}\aqu{1}+\qu{2}\aqu{2}+\qu{3}\aqu{3}+\qu{4}\aqu{4}-\quc{4}\aquc{4}-\quc{3}\aquc{3}-\quc{2}\aquc{2}-\quc{1}\aquc{1}\bigg).
\end{align}
In Fig. \eqref{fig:sigma_um_spectra}, we show the $\Sigma_u^-$ spectrum. Note that the slope of the energy levels are steeper than the Nambu-Goto spectrum, so we did not fit the ansatz of Eq. \eqref{eq:modified_nambu_goto} to the data. 

\section{Conclusions and outlook}
We succeeded to compute a significant number of excitations for different symmetries of the flux tube, improving the state of the art, Table. \eqref{Tab:number_excitations}. Considering a second parameter $\sigma_2$ in the Arvis potential, Eq. \eqref{eq:modified_nambu_goto}, results in better fits to the data. The values of $\sigma_2$ are almost $10\%$ smaller than $\sigma$, Fig. 
\eqref{fig:deviation_nambu}, leading to larger energy splitting between energy levels than the Nambu-Goto spectrum. This tension can be a signal for the existence of a constituent gluon in the excited flux tubes. As the $\chi^2/dof$ of the fits are large, we should take this deviation with a grain of salt.
\begin{minipage}{\textwidth}
	\centering
	\begin{minipage}{0.35\textwidth}
		\includegraphics[scale=0.9]{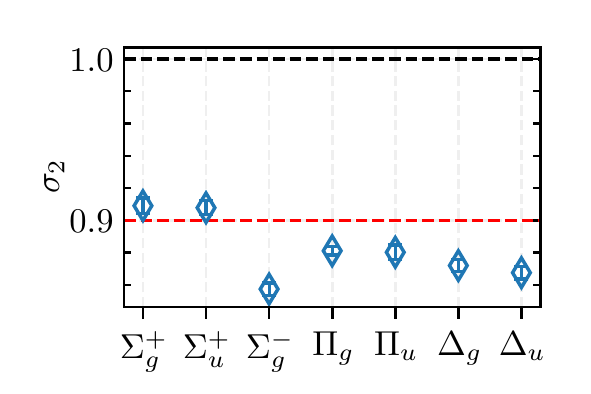}
		\captionof{figure}{Values of $\sigma_2$ for different symmetries of the flux tube.\label{fig:deviation_nambu}}
	\end{minipage}\hfill
	\begin{minipage}{0.6\textwidth}
		\centering
\begin{tabular}{c|c|c|c|c|c|c|c|c}
	\hline\hline	
\diagbox[width=5.5em]{Action}{	$\Lambda_\eta^+$}&	$\Sigma_g^+$&$\Sigma_g^-$&$\Sigma_u^+$&$\Sigma_u^-$&$\Pi_g$&$\Pi_u$&$\Delta_g$&$\Delta_u$\\
	\hline\hline
	$S_{II}$&2&0&0&0&1&1&1&0\\
	\rowcolor{gray}&8&6&4&2&6&6&6&4\\
	\hline\hline
\end{tabular}
	\captionof{table}{The first row shows the numbers of excitations are already reported in the literature \cite{Juge_2003}, the second row show our results.\label{Tab:number_excitations}}
\end{minipage}\hfil
\end{minipage}

\vspace{0.5 cm}

\section{Acknowledgments}
\textcolor{black}{ 
We acknowledge the discussion on flux tubes and our results with our colleagues Andreas Athenodorou, Bastian Brandt, Kate Clark, Caroline Riehl, Lasse Müller, and Marc Wagner.}
	Alireza Sharifian and Nuno Cardoso  are supported by FCT under the Contract No. SFRH/PD/BD/135189/2017 and SFRH/BPD/109443/2015, respectively. The authors thank CeFEMA, an IST research unit whose activities are partially funded by
FCT contract  UIDB/04540/2020 for R\&D Units.

\bibliography{references}
\bibliographystyle{unsrt}

\end{document}